\documentclass[jcp,aip,graphicx,,amsmath,amssymb,floatfix,reprint,citeautoscript, longbibliography]{revtex4-1}
\usepackage{graphics}
\usepackage{graphicx}
\usepackage[colorlinks,allcolors=black,citecolor=blue,urlcolor=blue]{hyperref}
\usepackage{upgreek}
\graphicspath{{figures/}}

\begin{document}

\title{%
Deciphering
High-order Structural Correlations within
Fluxional Molecules from Classical and Quantum
Configurational Entropy%
} 

\author{Rafa{\l} Topolnicki}
\email{rafal.topolnicki@uwr.edu.pl}
\affiliation{Lehrstuhl f{\"u}r Theoretische Chemie, Ruhr-Universit{\"a}t Bochum, 44780 Bochum, Germany}
\affiliation{Institute of Experimental Physics, University of Wroc{\l}aw, 50-204 Wroc{\l}aw, Poland}

\author{Fabien Brieuc}
\affiliation{Lehrstuhl f{\"u}r Theoretische Chemie, Ruhr-Universit{\"a}t Bochum, 44780 Bochum, Germany}

\author{Christoph Schran}
\affiliation{Lehrstuhl f{\"u}r Theoretische Chemie, Ruhr-Universit{\"a}t Bochum, 44780 Bochum, Germany}
\affiliation{Present Address: Department of Physics and Astronomy, University College London, London, WC1E 6BT, UK}

\author{Dominik Marx}
\affiliation{Lehrstuhl f{\"u}r Theoretische Chemie, Ruhr-Universit{\"a}t Bochum, 44780 Bochum, Germany}

\date{\today}

\begin{abstract}
We employ the $k$-th nearest-neighbor 
estimator of configurational entropy in order to decode
within a parameter-free numerical approach 
the complex high-order structural correlations in 
fluxional molecules going much beyond the usual linear, 
bivariate correlations.
This generic entropy-based scheme for determining
many-body correlations is applied to the complex configurational
ensemble of protonated acetylene, a prototype for
fluxional molecules featuring large-amplitude motion.
After revealing the importance of high-order correlations
beyond the simple two-coordinate picture for this
molecule, we analyze in detail the evolution of 
the relevant correlations with temperature
as well as the impact of nuclear quantum effects 
down to the ultra-low temperature regime 
of 1~K. 
We find that quantum delocalization and zero-point
vibrations significantly reduce all correlations
in protonated acetylene
in the deep quantum regime.
Even at low temperatures up to about 100\,K, 
most correlations are 
essentially absent in the quantum case and only gain importance 
at higher temperatures. 
In the high temperature regime, 
beyond roughly 800\,K, the increasing thermal fluctuations 
are found to exert a destructive effect on the presence of correlations. 
At intermediate temperatures of approximately 100 to~800\,K,
a quantum-to-classical cross-over regime is found
where classical mechanics starts to correctly describe
trends in the correlations whereas it even qualitatively 
fails below 100\,K.
Finally, a classical description of the nuclei provides 
correlations that are in quantitative agreement with the 
quantum ones only at temperatures exceeding 1000\,K.
This data-intensive analysis has been made possible due to
recent developments of machine learning techniques
based on high-dimensional neural network potential energy surfaces
in full dimensionality that allow 
us to exhaustively sample both, the classical and quantum
ensemble of protonated acetylene 
at essentially converged coupled cluster accuracy
from~1~to more than 1000\,K. 
The presented non-parametric
analysis of correlations beyond usual
linear two-coordinate terms is 
transferable to other system classes. 
The technique is also expected to complement 
and guide the analysis of experimental measurements,
in particular multi-dimensional vibrational spectroscopy,
by revealing the complex coupling between various degrees of freedom.
\end{abstract}

\maketitle 
\section{\label{sec:intro}Introduction}
As known for decades, the static and dynamical
properties of any chemical system are governed
by its potential energy surface (PES) within the
Born-Oppenheimer approximation.
In principle, it would therefore be sufficient
to sample the PES of the system of interest to
understand its properties.
However, it can very easily become a daunting task
to analyze in detail the molecular motion 
in cases where the involved coordinates feature non-trivial 
high-order correlations~--even if these systems would 
typically classified to be ``small'' since built from 
only a handfull of atoms. 
In such cases the analysis needs to go beyond commonly used
correlation coefficients that are only able to reveal
linear bivariate correlations.

In this study, we employ  a general framework to
decipher high-order correlations between any desired degrees of
freedom of a chemical system using concepts originating 
in information theory that are based on a non-parametric 
entropy estimator.
Here, the well-known
$k$-th nearest-neighbor configurational
entropy estimator \cite{Kozachenko1987, Singh2003, Mnatsakanov2008}
is employed to analyze the configurational ensemble obtained from 
sampling the PES, using molecular dynamics simulations.
Since this approach is very general and does not require 
any assumptions regarding topological properties of the
sampled configuration space, it has a broad range of 
applications.
In the context of molecular science, the method has previously been successfully
used for instance to estimate translational and orientational entropies
of small molecular systems~\cite{Huggins2014}.
In addition, it has been shown that this method, although suffering from slower
convergence in higher dimensions, can be applied to approximate the total
configurational entropy by a 
truncated mutual-information expansion~\cite{Matsuda2000, Killian2007, Hnizdo2008}.
Due to increasing computational resources, the $k$-th nearest-neighbor approach
has been recently employed in the field of moderately sized biomolecular systems,
where it supersedes much simpler and less accurate
parametric methods, such as the quasi-harmonic approximation~\cite{Karplus1981}
or Schlitter's entropy formula~\cite{Schlitter1993}.
These applications include, among others, the investigation of the
changes in configurational entropy due to binding and interactions 
of biomolecular systems with solvents~\cite{Fenley2014, Fogolari2015, Huggins2015}.
For comprehensive background and overview in the realm of chemistry, 
we refer the reader to a recent review article~\cite{Suarez2015} 
of both, parametric as well as non-parametric methods for entropy 
estimation from molecular simulations.

After presenting the required methodology for the analysis
of high-order correlations based on the configurational entropy,
we apply this new approach to the study of correlations between 
the coordinates of protonated acetylene as a function of temperature.
Protonated acetylene is a fluxional (or floppy) molecule
subject to large-amplitude motion, which can be activated
both by temperature or quantum effects~\cite{Marx1996}
and thus is a prime candidate for complex, non-linear correlations.
Although being a relatively small molecule, protonated acetylene 
is highly relevant for gas phase chemistry~\cite{Knoll2003,
Psciuk2007, Douberly2008, Fortenberry2016} and has a complex PES that offers 
the opportunity to analyze intricate internal motion.
Protonated acetylene is studied since decades by now in experiment and theory. 
In the beginning, the debate was focused on
predicting the correct global minimum energy structure.
Early Hartree-Fock electronic structure calculations 
suggested a Y-shaped or 'classical' isomer to be the energetically 
most favorable configuration~\cite{Hariharan1972}, while taking 
electron correlation into account yielded 
the bridge-shaped or 'non-classical' structure to be more stable~\cite{Zurawski1973}
as unveiled using pioneering coupled-cluster-like methods;
see Fig.~\ref{fig:neb} for visual representations of the two isomers.
It is now well established that the 
global minimum on the PES is the non-classical 
bridged isomer, while the classical structure is 
a shallow local minimum that is 
something like 17--21~kJ/mol
higher in energy~\cite{Psciuk2007}.
Later, a great share of attention focused on dynamical properties of the 
C$_2$H$_3^+$ molecule initiated by the results of Coulomb explosion imaging 
(CEI) 
experiments~\cite{Vager1993}, which has been further extended using 
combined ab initio and Monte Carlo techniques~\cite{Knoll2003}. 

A crucial component of any data-intensive correlation analysis 
such as the one to be utilized
in what follows
is the efficient sampling of the PES of 
the molecular system of interest using molecular simulations. 
In the present context of protonated acetylene,
this is made possible by the use of a machine learning approach 
based on the automated development of the so-called high-dimensional
neural network potentials (NNPs)~\cite{Behler2007, Behler2014, Behler2015, Behler2017}.
Here, we generate the global NN-PES that describes
the large-amplitude conformational dynamics of bridged and Y-shaped structures
at essentially converged coupled cluster accuracy~--
thus going beyond the electronic structure methods previously used
in this context~--
using a largely automated fitting procedure~\cite{Schran2020}.
This allows us to conduct an accurate statistical
sampling followed by exhaustive configurational entropy 
analysis based on extensive molecular dynamics simulations.
Notably, we describe the nuclei as both, classical and quantum
degrees of freedom,  
the latter obtained from converged path integral simulations,
from ultra-low to very high temperatures
without reducing the dimensionality of the problem. 
It therefore becomes possible to quantify the correlations
in protonated acetylene from both, the classical
and quantum configurational entropy and to study
their evolution as a function of temperature. 
This detailed analysis spans vastly different
conditions, from ultra-low temperatures of 1\,K
to ambient conditions up to very high temperatures
of 1600\,K and reveals the distinct differences
between a quantum and classical description of the nuclei
depending on the temperature regime.

The present  study showcases the great potential of 
our non-parametric 
correlation analysis approach based on configurational entropy to
address not only bivariate correlations, but also
to study high-order correlations, 
here up to four-point correlations in a fluxional molecule, 
which are expected to be of relevance much beyond the present case.
%

\section{\label{sec:methodology}Methodology}
%
\subsection{Entropy and interaction information}
%
The configurational entropy $S$ associated with coordinates
${\bf q}=(q_1,\ldots,q_s)$ of a single molecule is given by
\begin{equation}
S(1,2,\ldots,s) = -k_{\rm B} \int f({\bf q})\> \ln f({\bf q}) \> d{\bf q}
\enspace ,
\label{eqn:def_entropy}
\end{equation}
where $k_{\rm B}$ is the Boltzmann constant and $f({\bf q})$ is the
continuous probability density function of the coordinates used
to describe the molecular configurations.
In the classical case, this probability density 
simply is the classical Boltzmann distribution 
in ${\bf q}$-space, while for quantum systems
at finite temperatures, $f({\bf q})$ is given by 
the diagonal part of the thermal density matrix. 
For simplicity, we shall use $k_{\rm B}=1$ and express the entropy as a 
unitless quantity broadly known as information-theoretic or
Shannon entropy~\cite{Cover2006} 
and von~Neumann entropy in the context of quantum 
mechanics~\cite{Ohya1993,Henderson2001}. 
In information theory, the mutual information concept is usually used to describe
the correlation between variables.
The so-called mutual information $I(i,j)$ between the two variables 
$q_i$ and $q_j$ is defined as~\cite{Cover2006} 
\begin{equation}
 I(i,j) = S(i) + S(j) - S(i,j)
\enspace , 
\label{eqn:def_MI}
\end{equation}
where $S(i), S(j)$ and $S(i,j)$ are the entropies of the
systems~$i$, $j$ and $\{i,j\}$, respectively.
The function $I(i,j)$ measures the amount of information
about variable $q_i$ that is gained from a measurement
of variable $q_j$ and vice versa.
In other words, the mutual information $I(i,j)$ represents
the reduction of uncertainty about $q_i$ due to the knowledge of $q_j$.
Therefore, $I(i,j)$ quantifies the degree of correlation
between $q_i$ and $q_j$, i.e. the smaller the value
of the mutual information the more independent the two variables are.
%

A most widely used measure of correlation, in particular 
in the context of computational entropy estimation based 
on (bio)molecular simulations, is the correlation coefficient
matrix with elements 
$$ \rho_{i,j} = \frac{\mathrm{cov}(i, j)}{\sigma_i\sigma_j}, $$
where $\mathrm{cov}(i, j)$ is 
the  covariance between
variables $q_i$ and $q_j$; $\sigma_i$ and $\sigma_j$
are the usual standard deviations of these variables.
We note in passing that this is also the general idea underlying
what is called principal component analysis (PCA),
principal mode analysis (PCM), or 
essential dynamics (ED) depending on the community. 
In stark contrast to $\rho_{i,j}$, which is only sensitive 
to linear correlations, the mutual information $I(i,j)$
characterizes a general dependence and, thus, is able to
perfectly quantify also non-linear correlations among the considered
variables~\cite{Smith2015}.
Moreover, the mutual information is invariant under invertible 
transformations of the data as opposed to the correlation coefficient.
This property is especially desirable when detecting the correlated 
motion within fluxional molecules since no assumptions
about the nature of the correlation are required.
Therefore, investigating the mutual information provides
a very general framework that may be used to study all
kinds of correlation among a suitable set of generalized 
coordinates describing the arrangement of particles 
(such as atoms or nuclei) in space. 

The mutual information defined above can be generalized to
describe higher-order correlations.
One of possible approaches for such a generalization is
the so-called interaction information~\cite{McGill1954}.
For a given subset $U=\{i_1,\ldots,i_n\}\subseteq \{i_1,\ldots,i_s \}$ 
of $n$ variables the $n$-coordinate interaction information is
defined as~\cite{Timme2014}
\begin{equation}
I_n(U) = - \sum_{T \subseteq U}(-1)^{|U|-|T|}S(T)
\enspace ,
\label{eqn:def_II}
\end{equation}
where the sum runs over all possible subsets 
$T \subseteq U$ and $|U|$ denotes the set size of $U$.
It can be seen that the first-order interaction information $I_{1}(i)$
is the entropy $S(i)$ itself, the mutual information Eq.~(\ref{eqn:def_MI})
is a special case of interaction information Eq.~(\ref{eqn:def_II})
for $n=2$, which is the reason why we will use the term interaction information 
also in context of two-body interactions represented by mutual information.
The three-coordinate interaction information can be easily derived
from Eq.~(\ref{eqn:def_II}) to be
\begin{eqnarray*}
I_3(i,j,k)  & = & -S(i)-S(j)-S(k) \\
& + & S(i,j)+S(j,k)+S(i,k) \\
& - & S(i,j,k) \enspace.
\end{eqnarray*}
In general, the $n$-coordinate interaction information $I_n$ measures 
that contribution to the intrinsic correlation between $n$ 
coordinates~\cite{Matsuda2000} which is not already described by 
any of the lower-order correlations,
i.e.
$I_{n-1},I_{n-2},\ldots,I_2$.
In other words: 
The interaction information can also be viewed as the amount of
information that is common to all the attributes, but not present
in any subset~\cite{Timme2014}. 

Note that, while the two-coordinate interaction information $I_2$
is a non-negative quantity, the generalized $n$-coordinate
interaction information $I_n$, $n\ge 3$, can be both positive as
well as negative.
The positive interaction information is commonly referred to
as "synergy" as it implies the synergistic interaction
between the variables involved: We obtain more information
about the system by observing $n$ variables simultaneously
than we would obtain knowing all subsets containing at most
$n-1$ variables.
On the other hand, the negative interaction information 
implies a redundant interaction among the variables~\cite{Timme2014}.
Similar to mutual information, the interaction information
allows one to detect and to quantitatively characterize any kind of
correlation including non-linear components.
%

It should be noted that the approach presented above to study correlations
among the coordinates used to describe the arrangement of a molecule does not
depend on the actual coordinates that have been used.
Although the value of the configurational entropy depends on the chosen
coordinate system~\cite{Hnizdo2010}, the interaction information
does not.
The choice of a particular coordinate system,
meaning the set of generalized variables ${\bf q}$ used to compute $I_n$, is 
mainly suggested by its useful geometrical interpretation for the given problem
such as depicted below for protonated acetylene, see Fig.~\ref{fig:coordinates}.
The introduced methodological framework is therefore
the ideal approach to study very general correlations of
complex molecular motion.

\subsection{Entropy estimation}
%
In this work, we aim at quantifying correlations within fluxional
molecules using the interaction information which requires the estimation
of the configurational entropy. 
The configurational entropy $S(1,\ldots,s)$ of a general probability
distribution $f(q_1,\ldots,q_s)$ can be estimated from $N$ observations
${\bf x}^{(i)}=(x_1^{(i)},\ldots,x_s^{(i)}), i=1,\ldots,N$ of a
random vector ${\bf q}=(q_1,\ldots,q_s)$ 
(which will be the set of generalized coordinates defined in Fig.~\ref{fig:coordinates}
as generated from classical and quantum simulations of protonated acetylene)
based on the nearest-neighbor
distances between sample points.
The asymptotically unbiased and consistent estimator~\cite{Singh2003}
of the entropy is given by
\begin{equation}
S_k^{(N)} = \frac{1}{N}\sum_{i=1}^N \ln R_{i,k} + \ln \frac{N\pi^{s/2}}{\Gamma(\frac{s}{2}+1)} - L_{k-1}+\gamma
\enspace ,
\label{eqn:knn_entropy}
\end{equation}
where $R_{i,k}$ is the distance between the sample point ${\bf x}^{(i)}$
and its $k$-th nearest neighbor in the sample.
The last two terms on the right side of Eq.~(\ref{eqn:knn_entropy}),
namely $L_k=\sum_{j=1}^k 1/j, j\ge 1, L_0=0$ 
as well as the Euler--Mascheroni constant $\gamma$, are introduced 
to provide a bias correction~\cite{Singh2003}.
In the following, we will refer to the estimator according to 
Eq.~(\ref{eqn:knn_entropy})
as the $k$-th~NN estimator of the configurational entropy $S$;
keep in mind that this ``NN'' in the context of estimators
does not encode the notion ``neural network''. 
%

Very importantly, this 
$k$-th~NN entropy estimator is a non-parametric estimator
and therefore requires no assumptions about the functional form
of the underlying probability density function $f$,
in particular not that of a multivariate Gaussian as traditionally
assumed  in the context of computational entropy estimates based
on (bio)molecular simulations.
Moreover, the $k$-th~NN estimator has several highly desirable
properties such as being adaptive, data efficient and having
the minimal bias at given finite sample size $N$
(see Ref.~\citenum{Kraskov2004}).
The main drawback is the high computational complexity of the NN
searching algorithms which increases significantly with the 
dimensionality of the data.

One assumption that is made to derive Eq.~(\ref{eqn:knn_entropy})
is that the underlying probability density $f$ is constant in the
region of $k$ nearest neighbors around each sample 
point~\cite{Kozachenko1987, Kraskov2004}.
This assumption is better fulfilled for small values of $k$,
therefore it is widely accepted to use
$k=1,\ldots,5$
(see e.g. Refs.~\citenum{Hnizdo2007, Hnizdo2008}).
On the other hand, the parameter $k$ can be viewed as
a smoothing parameter where large values of $k$
corresponds to smoother estimates of the underlying
probability density function $f$, thus providing
a lower variance at the price of a larger bias, 
whereas using small values of $k$ provide smaller biases but
larger variances~\cite{Duda2001, Kung2012}.
Yet, in the limit $N \rightarrow \infty$, all $k$-NN estimators
should yield the same result regardless of the value of the
parameter $k$.
The asymptotic properties of the $S_k^{(N)}$ estimator were
derived long ago~\cite{Singh2003}:
The asymptomatic variance of the $k$-NN entropy estimator decreases with $k$.
However, for finite sample sizes $N$, as implied in practice 
by any numerical sampling of ${\bf q}=(q_1,\ldots,q_s)$ based 
on molecular simulations, the interplay between bias
and variance remains unknown.
It is also not clear how the errors in entropy estimation
transfer to errors in estimation of interaction information.
%

In order to cope with this practical limitation, it has been proposed in 
the literature~\cite{Hnizdo2007} to extrapolate
the entropy estimate to the limit $N\rightarrow\infty$ based on values
obtained on several data sets of increasing size.
This approach is time consuming as the estimation procedure
needs to be performed multiple times but also a specific form
of extrapolation function has to be imposed.
Since the true value of the entropy remains unknown, we are
unable to determine the bias of the $k$-NN estimator in the 
present case.
However, the variance can be approximated using the bootstrap 
technique~\cite{Efron1994}.
It turns out that the variance of the interaction information
Eq.~(\ref{eqn:def_II}) estimator rapidly decreases with the
few first values of $k$ and then remains almost constant as
$k$ increases further as explicitly shown in Sec.~III.C of the 
Supplementary Information.
Therefore, it is reasonable not to use the lowest possible
value of $k$ as the obtained estimate will be prone to
stochastic fluctuations.
Since the estimated value of the bias is unknown, we propose
here to use a rather small value of $k=5$ in order to benefit
from the rapid decay of the variance without introducing
significant systematic errors.
In order to validate our choice we have considered
values of $k$ up to 50 and virtually no quantitative
difference was observed compared to $k=5$ as discussed in detail
in Sec.~III.C of the Supplementary Information.

In the remainder of this paper, we are going to use the entropy estimator 
as defined via Eq.~(\ref{eqn:knn_entropy})
in order to construct a plug-in estimator of the 
$n$-point interaction information according to Eq.~(\ref{eqn:def_II}).
As will be demonstrated for protonated acetylene, 
the above-described estimation of the configurational
entropy based on the $k$-NN entropy estimator
enables the reliable analysis of general, high-order
correlations beyond using standard correlation coefficients
and parametric estimators. 
Computing $n$-coordinate interaction information based on such classical 
and quantum entropies provides a unique framework for the investigation 
of both, complex classical and quantum 
molecular motion as present in fluxional molecules.

\section{Computational Details}
\label{sec:comp_det}
%
The potential energy surface of the protonated acetylene molecule
that covers interconversion of its bridged and Y-shaped isomers 
as exposed in the introductory paragraphs 
is described using a high-dimensional NNP
\cite{Behler2007, Behler2014, Behler2015, Behler2017}.
This global NN-PES 
(dubbed ``V1-PES-Protonated-Acetylene-2020'') 
has been parameterized by us using our in-house
\texttt{RubNNet4MD} neural network package
\cite{RubNNet4MD}
based on a total of about 28~000~configurations
for which the energy has been computed using 
CCSD(T*)-F12a/aug-cc-pVTZ electronic structure calculations
~\cite{Adler2007, Knizia2009} with consistent scaling of 
the perturbative triples~\cite{Knizia2009}
(denoted for brevity throughout as CCSD(T*) in what follows) 
all performed using the \texttt{Molpro} package~\cite{Molpro}.
This particular explicitly correlated electronic structure 
method provides coupled cluster energies close to
the complete basis set (CBS) limit~\cite{Knizia2009}. 
Using this method to compute the reference data used to fit the present NN-PES
allows us to perform long and stable classical and path integral quantum
molecular dynamics simulations
of protonated acetylene at essentially converged 
coupled cluster accuracy.
In order to sample the configuration space of that fluxional molecule efficiently
and to keep the number of the computationally demanding
coupled cluster reference calculations as low 
as possible we applied our automated fitting scheme 
as introduced recently~\cite{Schran2020}.
Using 28~000 reference calculations, the
root-mean-square error (RMSE) in the training set is around 0.03\,kJ/mol
per atom, which corresponds to 0.15\,kJ/mol, thus being much better than 
the usually accepted ``chemical accuracy'' (i.e. 1\,kcal/mol or about 4\,kJ/mol) 
which indicates the high quality of our global NN-PES fit. 
We refer the interested reader to Sec.~I of the Supplementary Information for 
all details on the NNP architecture and the fitting procedure.

\begin{figure}
\centering{}
\includegraphics[width=\linewidth]{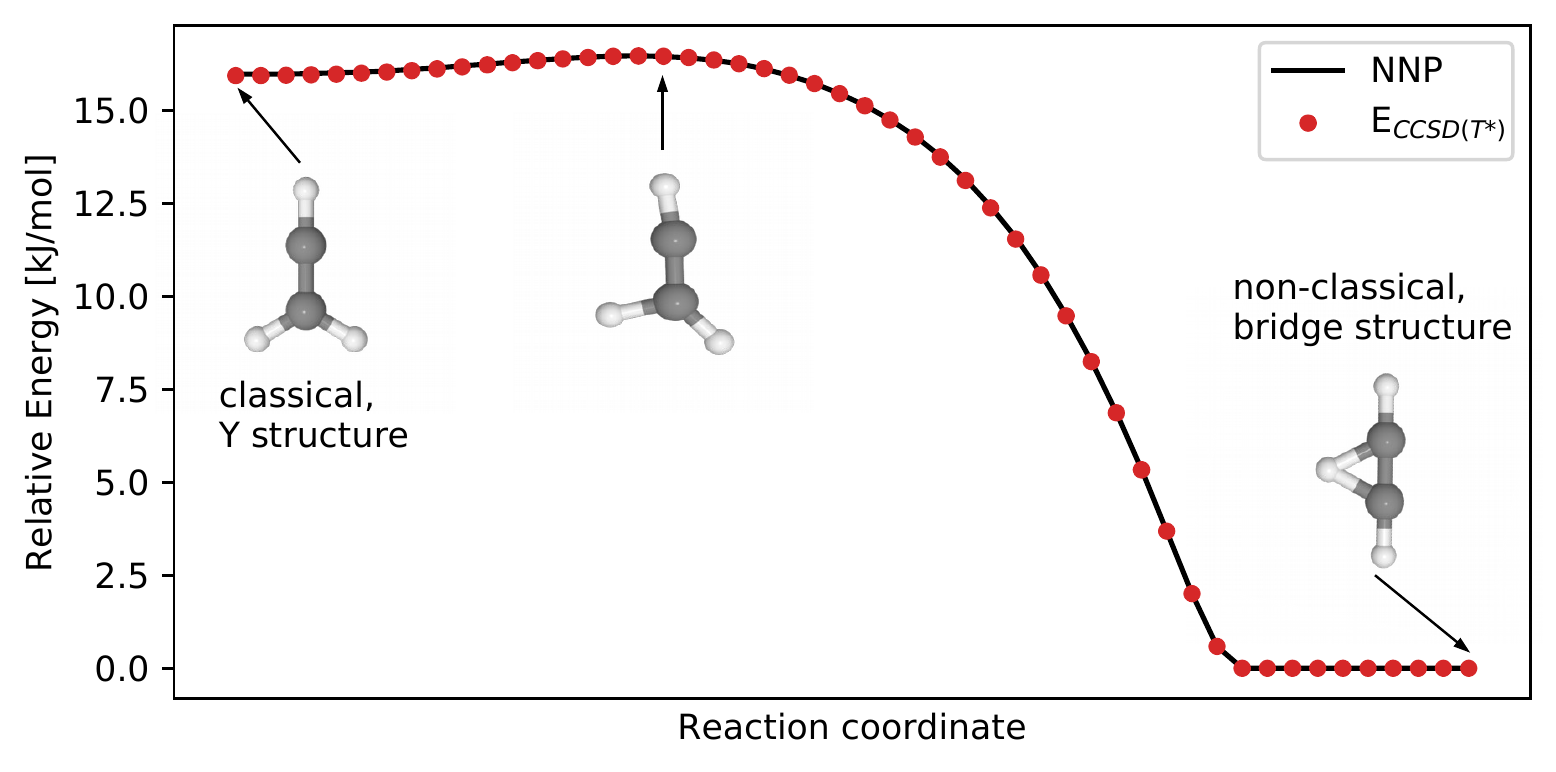}
\caption{
Energy profile along the minimum energy path (solid black line) 
as given by the global NN-PES of CCSD(T*) accuracy, see text. 
Along this path, single-point energies obtained by
using the identical electronic structure method are shown 
as red dots for direct comparison.
The global and local minimum of the C$_2$H$_3^+$ molecule
as well as the interconnecting saddle-point
corresponding to the bridge, Y-shaped and transition state structures
are depicted using ball-and-stick representations
in the right, left and middle parts of the figure, respectively. 
}
\label{fig:neb}
\end{figure}

Multiple tests were performed to validate the accuracy of this NN-PES
for protonated acetylene.
This includes, among others discussed in the Supplementary Information, 
the calculation of the transition path between the bridged and Y-shaped
conformations (using the improved tangent nudged elastic
band method (NEB) method~\cite{Henkelman2000}) obtained from this NN-PES 
as presented in Fig.~\ref{fig:neb}.
Along this minimum energy path, a dense set of single-point energies was computed 
for comparison using the same CCSD(T*) methodology. 
The NN-PES is able to reproduce not only the 16.46\,kJ/mol energy barrier
of the rearrangement from the Y-shaped to the bridged isomer with essentially
perfect agreement to the coupled cluster reference,
but also the overall shape of the transition path 
including the extremely shallow local minimum of the Y-structure, 
such that the naked eye cannot recognize any difference
on the intrinsic energy scale of that PES;
we note in passing that this important energy pathway is
consistent with the results reported earlier~\cite{Psciuk2007}.
For comprehensive benchmarking, we refer to data obtained 
by evaluating the NN-PES in direct comparison to explicit 
single-point CCSD(T*) calculations
(including energy predictions across the data set,
important stationary-point energies, 
normal modes of the key minimum-energy structures,
as well as potential energy scans along various internal generalized coordinates),
which are compiled in Secs.~I.A to~D of the Supplementary Information.
Overall, these tests confirm that the NN-PES fit is able to reproduce 
the coupled cluster PES of protonated acetylene very accurately.

This NN-PES was used to perform classical molecular dynamics (MD) and
quantum path integral molecular dynamics (PIMD) simulations
employing the \texttt{CP2k} simulation package~\cite{cp2k, Hutter2014}.
In this context, we refer the interested reader
to a recent review article~\cite{Brieuc2020}
that unfolds the entire methodological framework that allows one
to carry out converged quantum simulations of fluxional molecules or 
complexes even at cryochemical conditions. 
For the production runs, protonated acetylene 
was simulated at different temperatures ranging from 1\,K to 1600\,K 
using both, MD and PIMD simulations. 
In case of the PIMD simulations, the so-called PIQTB~\cite{Brieuc2016}
thermostat as implemented by us in \texttt{CP2k} for usage down to ultra-low 
temperatures~\cite{Schran2018} was utilized in order to converge 
the path integral in terms of its discretization.
This Trotter convergence was examined in detail for simulations
at 100\,K and compared with the corresponding
results obtained with standard canonical PIMD simulations using the
so-called PILE~\cite{Ceriotti2010} thermostat and a very large number
of path integral replica.
It was observed that in case of PIQTB using $P=48$ replica at 100\,K is
sufficient to provide converged results in terms of energetic
as well as structural properties; see Sec.~II in the
Supplementary Information for details.
It was also verified that both thermostats provide the same
values of the quantity of interest, namely the interaction information
$I_n$ up to 4-point correlations, as discussed in detail in Sec.~II.C of 
the Supplementary Information.
The number of replicas $P$ used for the simulations at
different temperatures was determined 
as usual in such a way that the product
$P \cdot T$ remains constant, thus providing similar 
relative discretizations at all temperatures,
thus using $P=4\>800$ replica at our ultra-low temperature of 1\,K.
At very low temperatures, this
approach was further validated by explicit benchmark simulations at
5\,K where taking $P=960$ was shown to indeed provide converged results;
we refer to Secs.~II.A to~B in the Supplementary Information for more details.
All reported simulations were propagated for at least 2\,ns in
the case of PIMD simulations and at least for 82\,ns for classical MD simulations. 
Overall, this adds up to a grand total of 2.2~$\upmu$s of
classical and quantum simulations of protonated acetylene 
carried out at CCSD(T*) accuracy. 
A time step of 0.25\,fs was used throughout and the first 2.5\,ps 
of each simulation was discarded to account for thermalization. 

\begin{figure}[tb]
\centering{}
\includegraphics[width=\linewidth]{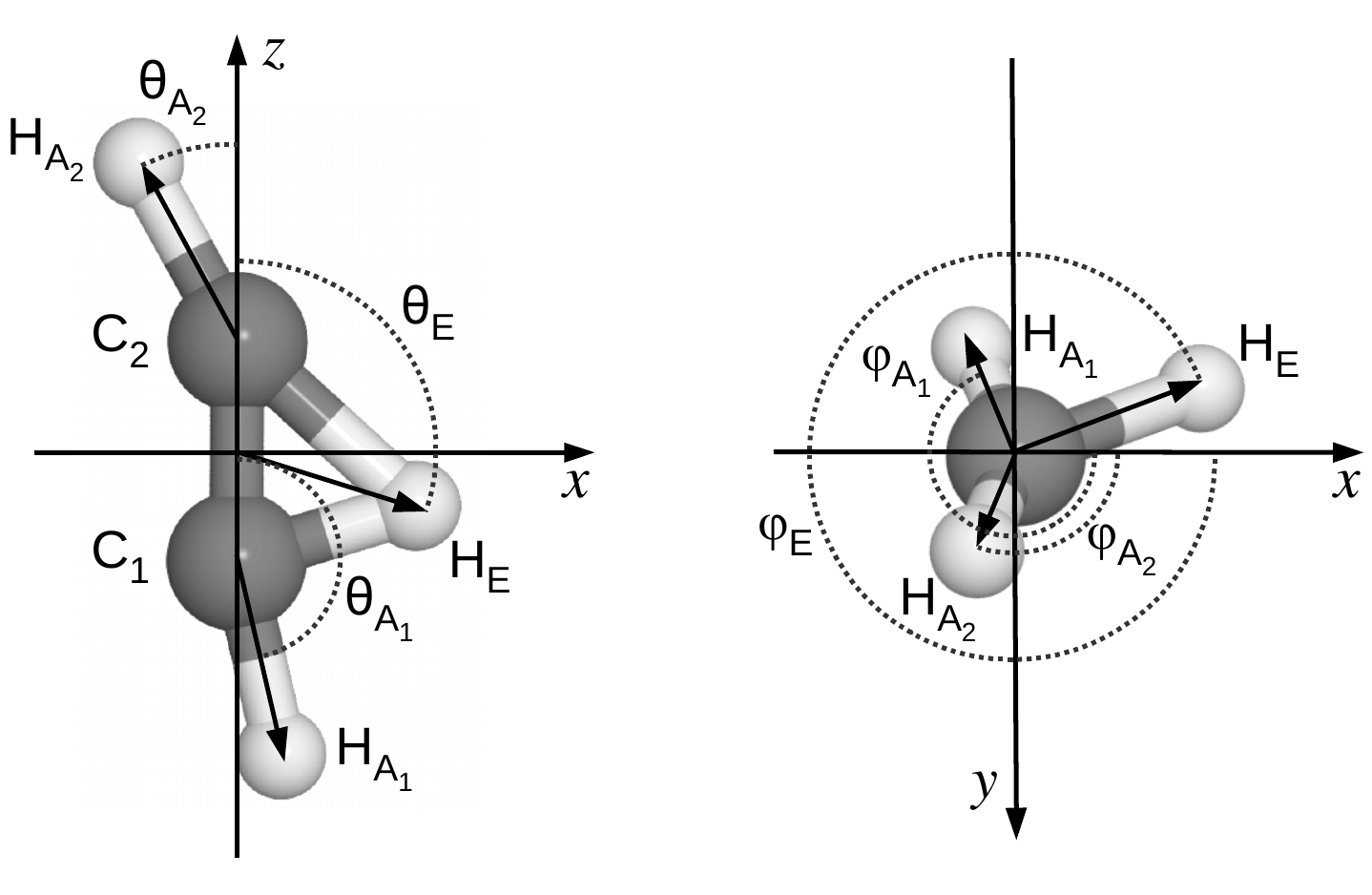} 
\caption{
Definition of polar angles $(\theta_{\rm A_1}, \theta_{\rm A_2}, \theta_{\rm E})$, left panel,
and azimuth angles $(\phi_{\rm A_1}, \phi_{\rm A_2}, \phi_{\rm E})$, right panel,  used to
describe 
the orientational configuration of the three protons with respect to the carbon atoms of the C$_2$H$_3^+$ molecule.
The corresponding distances of the axial and equatorial protons 
($r_{\rm A_1}, r_{\rm A_2}, r_{\rm E}$) are not labeled in the
figure but defined therein using arrows. 
All coordinates are defined with respect to 
a laboratory-fixed coordinate system as follows:
The molecule is translated and rotated with respect to the $x$- and 
$z$-axes such that the two C~atoms lie on the $z$-axis 
and that the origin is given by the C---C bond midpoint.
}
\label{fig:coordinates}
\end{figure}

The structural correlation analysis was performed using a set of
ten generalized coordinates of C$_2$H$_3^+$, which are the C---C bond length
($r_{\rm CC}$), the C---H bond lengths of the two axial ($r_{\rm A_1}, r_{\rm A_2}$) protons,
the distance between the equatorial proton and the C---C midpoint ($r_{\rm E}$),
the polar angles of each proton ($\theta_{\rm A_1}, \theta_{\rm A_2}$ and $\theta_{\rm E}$)
and the azimuth angles of each proton denoted as $\phi_{\rm A_1}, \phi_{\rm A_2}$
and $\phi_{\rm E}$, respectively. 
This set of coordinates allows one to uniquely define 
all possible configurations of the protonated acetylene molecule
within the laboratory-fixed coordinate system as defined in Fig.~\ref{fig:coordinates}
together with the graphical definition of all ten coordinates.
We note in passing that this set is very convenient for the
present discussion but slightly redundant since only nine internal 
degrees of freedom exist for C$_2$H$_3^+$.
In case of the azimuth angles of the protons, $(\phi_{\rm A_1}, \phi_{\rm A_2}, \phi_{\rm E})$,
periodic boundary conditions with a period of $2\pi$ apply.
As a consequence, the distance between points $\phi_i$ and $\phi_j$
is given by
$$
d_{\phi}(\phi_1, \phi_2) = \min\{|\phi_i-\phi_j|,  2\pi-|\phi_i-\phi_j| \},
$$
which can be easily generalized to higher dimensions.

By construction, our generalized coordinates are not invariant 
with respect to permutations,
e.g. exchanging atoms H$_{\rm A_1}$ with H$_{\rm E}$ changes the values
of $r_{\rm A_1}, r_{\rm E}$ and $\phi_{\rm A_1}, \phi_{\rm E}$.
In particular, these coordinates are only meaningful 
for bridge-like structures, see Fig.~\ref{fig:neb} for details. 
However, as evidenced by the minimum energy path in Fig.~\ref{fig:neb},
the local minimum of the Y-configuration is extremely shallow
with a barrier of only about 0.5~kJ/mol toward the global minimum,
whereas the reverse barrier of the global minimum toward the Y-structure
amounts to $\approx 16.5$~kJ/mol. 
Thus, the relative population of Y-like structures compared to 
bridge-like structures is expected to be overall very small at finite temperatures. 
In an effort to nevertheless take into account possible atom exchanges
and thus permutations of axial versus equatorial proton labels
when computing the respective generalized coordinates, 
we consider in each simulation step
all possible permutations of the three~H and two C~nuclei
(twelve permutations in total) according to that configuration
and define a structure in a given permutation state to be bridge-like if
the following set of conditions is fulfilled by the three polar angles:
$3\pi /4 \le \theta_{\rm A_1} \le \pi$,
$0 \le \theta_{\rm A_2} \le \pi /4$, and
$\pi/2-0.6 \le \theta_{\rm E} \le \pi/2+0.6$.
These threshold values used to classify a given structure
as bridge-like has been determined from
the respective minima of the distribution functions
of the polar angles when no permutations are applied 
as presented in Sec.~III.A of the Supplementary Information.
Modifying these threshold values 
in meaningful bounds only slightly impacts on
the quantitative results, 
whereas the qualitative findings presented below remain unchanged.
Indeed, explicit simulation shows that only very few structures, below 1~\%, are
in practice rejected by these criteria up to temperatures of 300\,K,
which is fully in line with the expected instability
of Y-like structures given the
shape of the interconversion profile Fig.~\ref{fig:neb}.

Standard Nos\'e-Hoover-chain thermostatted molecular dynamics 
is used here to sample the classical canonical phase space 
distribution function at a given temperature, and thus to 
generate molecular structures in configuration space which
 are distributed according to the classical Boltzmann distribution.
In the quantum case, path integral MD (PIMD) is correspondingly
used to generate the configurations according to the canonical
density matrix at the selected temperatures as detailed in 
Secs.~II and~III of the Supplementary Information.
The internal coordinates are then constructed from these
configurations and used as sample points for the computation 
of the associated entropies using the $k$-th NN estimator.
In the PIMD case, the atomic positions of all replica were treated 
independently and the corresponding coordinates, computed for 
each replica separately, were added to the data set utilized to compute 
the $k$-th NN estimator, thus following the usual procedure to compute
quantum expectation values of observables which are only 
defined in position space.

Prior to computing the $k$-th NN entropy estimator, values of all 
the coordinates, except the azimuth angles $\phi_{\rm A_2}, \phi_{\rm A_2}$ 
and $\phi_{\rm E}$, are standardized so that they have the same 
standard deviation as the azimuth coordinates. 
This is done by subtracting the mean, dividing by the standard 
deviation of the given coordinate, and multiplying by the average
standard deviation of azimuth coordinates.
As already said, the interaction information according to 
Eq.~(\ref{eqn:def_II}) is invariant under such linear transformations, 
but in this way the effective distance scale between points in 
different dimensions is more homogeneous.
This offers the numerical advantage that the convergence of the $k$-th NN
entropy estimators is faster in higher dimensions.

In order to determine the distances $R_{i,k}$ required to
compute the $k$-th nearest neighbor of all sample points, 
we used the ANN code~\cite{ANN} which utilizes the 
so-called ``kd~tree'' algorithm~\cite{Arya1993, Arya1998}.
Although with ANN it is possible to determine 
approximate nearest neighbors to reduce the computational complexity
we decided to use here an exact algorithm instead~\cite{ANN, Arya1993b}.

The interaction information Eq.~(\ref{eqn:def_II}) can be estimated directly 
from the data~\cite{Kraskov2004}, which usually provides
a faster convergence with respect to the number of data points at a price 
of higher computational complexity.
However, it turns out that, due to the high performance of
NN-PES molecular dynamics simulations, it is
computationally less demanding to generate large sets 
of uncorrelated structures and 
to
use the $k$-th NN entropy estimator 
Eq.~(\ref{eqn:knn_entropy}), combined with Eq.~(\ref{eqn:def_II}),
than to estimate the interaction information directly from the data.

\section{Results}
\label{sec:res}
%
In order to provide detailed insight into the correlations in fluxional
molecules based on the above introduced configurational entropy analysis,
we set out to study the temperature dependence of protonated acetylene,
a prime candidate of large-amplitude motion and fluxionality.
For that purpose, we performed exhaustive classical and quantum
simulations spanning vastly different regimes:
From ultra-low temperatures down to 1\,K to ambient conditions
to the highest temperature of 1600\,K.
The computational details of these simulations can be found
in Sec.~\ref{sec:comp_det} with further references to the
Supplementary Information.
Concerning the convergence of the interaction
information estimators with respect to number of data points
we refer to Sec.~III.B in the Supplementary Information.
The resulting classical and quantum ensembles at these distinct
temperatures enable the detailed analysis of the evolution
of correlations as a function of temperature, while
at the same time unraveling the importance of
nuclear quantum effects in different temperature regimes.

To start the detailed configurational entropy analysis
of correlations in protonated acetylene,
all possible two-coordinate interaction informations for the 
ten coordinates presented in Fig.~\ref{fig:coordinates} 
were determined from the classical and quantum trajectories
at several distinct temperatures.
As an important first result, we observe that there are only
three significant two-coordinate correlations.
All other correlations are several times weaker and
therefore will not be considered in the following.
A detailed discussion of all correlations can be found
in the Supplementary Information in Sec.~IV.A,
where we also show that the relative importance of the
correlations neither depends on the temperature nor
on treating the nuclei as classical or quantum point particles. 
The first prominent correlation, $I_2(r_{\rm E}, \theta_{\rm E})$, is
between the distance of the equatorial proton from the C---C bond,
$r_{\rm E}$, and its polar angle $\theta_{\rm E}$.
This is the only correlation that involves any of the existing
bond lengths with protonated methane, i.e. all other correlations 
considered here, including the 
three- and four-coordinate ones, 
exclusively involve angular degrees of freedom.
The second prominent correlation, 
$I_2(\theta_{\rm A_{1}},\theta_{\rm E})$ and its symmetry equivalent
$I_2(\theta_{\rm A_{2}},\theta_{\rm E})$, 
involves the polar angle of 
one of the two axial protons (i.e. $\theta_{\rm A_1}$ or $\theta_{\rm A_2}$) 
and the polar angle of the equatorial proton $\theta_{\rm E}$.
The last significant two-point correlation, 
$I_2(\phi_{\rm A_1}, \phi_{\rm A_2})$, is observed between the azimuth 
angles of the two axial protons.
Interestingly, the correlations between 
the azimuthal orientations 
of two axial protons is not mediated by the orbiting equatorial proton:
Correlations between the variable $\phi_{\rm A_1}$ and any of three
coordinates describing the position of the equatorial proton
(i.e. $r_{\rm E}, \theta_{\rm E}, \phi_{\rm E}$) are always at least one order of
magnitude weaker then $I_2(\phi_{\rm A_1}, \phi_{\rm A_2})$.
In fact, the interaction information between the azimuth angle of
axial and equatorial protons, which seems to be a natural
candidate to proxy the angular position of axial protons,
is virtually zero, which proves that $\phi_{\rm A_1}$ 
and $\phi_{\rm E}$ are independent.
In other words: Analysis of the two-point correlations suggests that
the equatorial proton orbits around the molecular axis given by 
the C--C bond independently from the orientational arrangement of 
the two axial protons. 
However, the higher-order correlations paint a different picture
as follows.

Let us next focus on the relevant higher-order correlations
in protonated acetylene, while as before discussion of
all correlations can be found in the SI in Sec.~IV.B.
There are only two significant three-coordinate correlations found
which are in addition equivalent by symmetry: 
$I_3(\theta_{\rm E}, \phi_{\rm A_1}, \phi_{\rm E})$ and $I_3(\theta_{\rm E}, \phi_{\rm A_2}, \phi_{\rm E})$.
They involve the two angular coordinates of the equatorial
proton, namely the polar $\theta_{\rm E}$ and azimuth $\phi_{\rm E}$ angles,
and the polar angle of one of the two axial protons, i.e.
either $\phi_{\rm A_{1}}$ or $\phi_{\rm A_{2}}$.
This result indicates that there exists a higher-order correlation between the
orientations of the protons that cannot be explained by 
the respective two-coordinate correlations discussed above. 
In particular, no substantial two-coordinate correlations involving
the $\phi_{\rm E}$ variable are recognized.
Therefore, it can be concluded that the position of the orbiting
equatorial proton is coupled in a non-trivial way to the arrangement 
of either one of the two axial protons and vice versa. 
Such a correlation has not been recognized previously~\cite{Marx1996, Knoll2003}
due to the lack of appropriate analysis methodology. 
Notably, this three-coordinate correlation is found to be
stronger than any of the two-coordinate correlations involving only
angular coordinates and, therefore, cannot be interpreted
as a minor correction only. 

Among all possible four-coordinate correlations there is only the
correlation given by $I_4(\theta_{\rm E}, \phi_{\rm A_1}, \phi_{\rm A_2}, \phi_{\rm E})$
which is not negligible.
However, the magnitude of $I_4$ is roughly half the value of $I_3$
and it features negative values in addition. 
This four-body correlation links the azimuth angles of all protons
with the polar angle of the equatorial proton, hence connects only
the coordinates that are present in the two three-coordinate correlations
discussed in previous paragraph.
This explains why the four-coordinate interaction information is negative.
According to Eq.~(\ref{eqn:def_II}), its negative value can be interpreted
as a redundancy in the information provided by the involved coordinates and
usually indicates a common-cause relation between the involved variables. 
In our case this indicates that the azimuth position of the $\phi_{\rm A_2}$ 
coordinate is already partially determined by the values of 
$\theta_{\rm E}, \phi_{\rm A_1}$ and $\phi_{\rm E}$ 
(an analogous relationship holds when $\phi_{\rm A_2}$ is replaced by $\phi_{\rm A_1}$
and vice versa).

\begin{figure}
\centering{}
\includegraphics[width=0.8\linewidth]{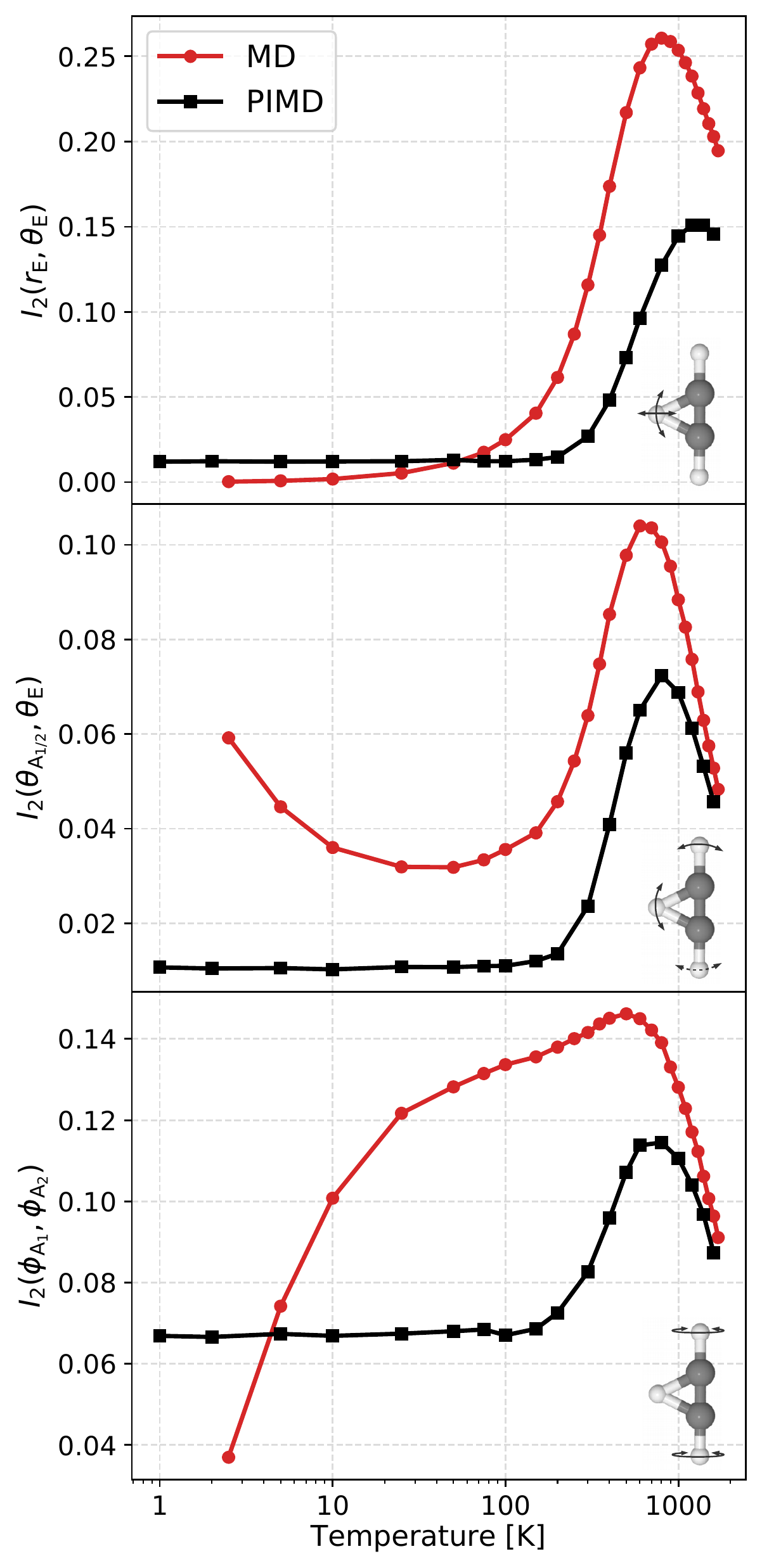}
\caption{
Significant two-coordinate correlations of protonated acetylene
as a function of temperature:
$I_2(r_{\rm E}, \theta_{\rm E})$, 
$I_2(\theta_{\rm A_{1}},\theta_{\rm E})$ and its symmetry equivalent
$I_2(\theta_{\rm A_{2}},\theta_{\rm E})$, as well as 
$I_2(\phi_{\rm A_1}, \phi_{\rm A_2})$
from top to bottom.
Correlations computed based on configurations generated
using classical molecular dynamics (MD) simulations are
marked using red lines, while the corresponding
quantum path integral molecular dynamics (PIMD) data
are given by black lines.
\label{fig:2body}
}
\end{figure}

\begin{figure}
\centering{}
\includegraphics[width=0.9\linewidth]{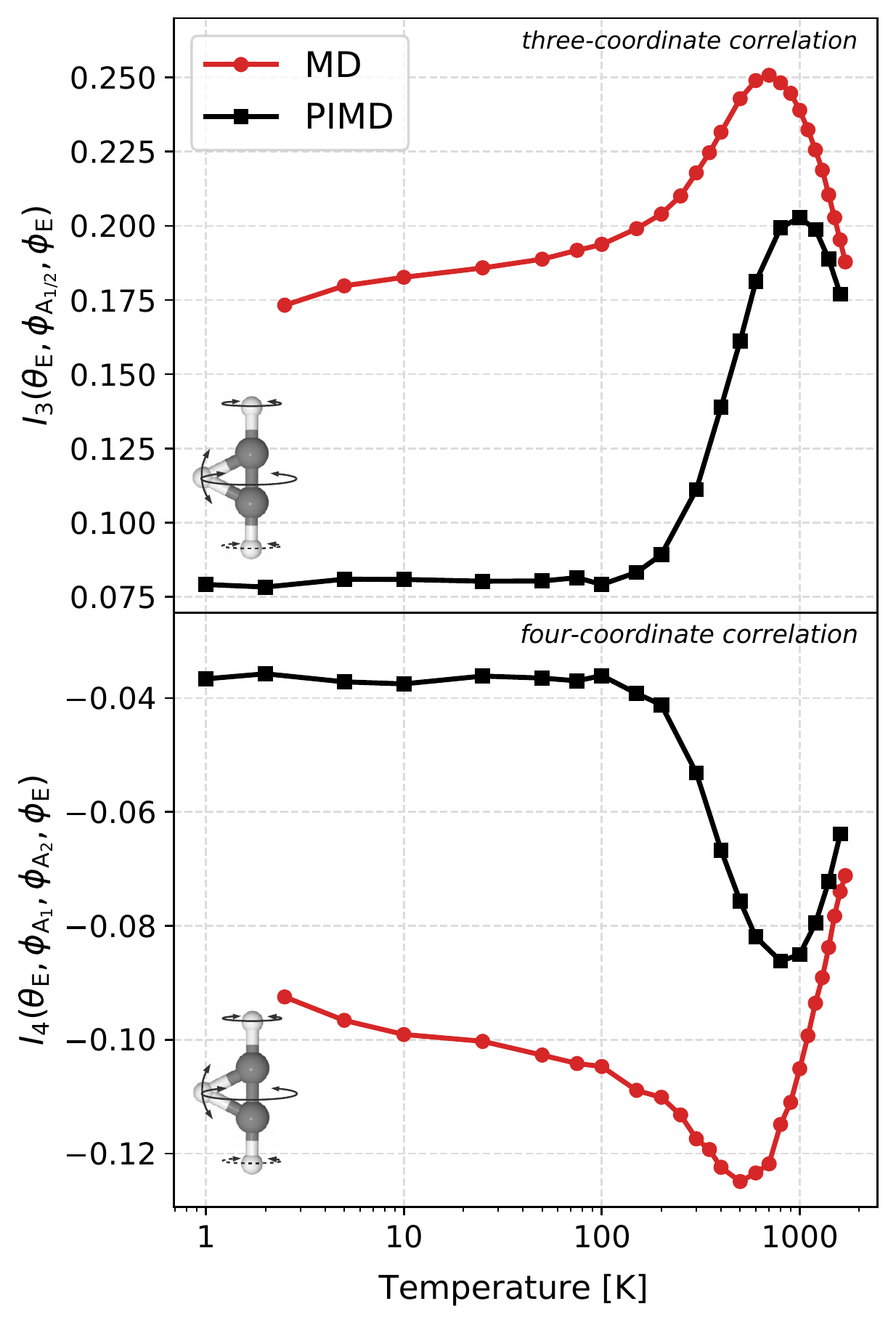}
\caption{
Significant three-coordinate (top) and four-coordinate (bottom)
correlations of protonated acetylene
as a function of temperature:
$I_3(\theta_{\rm E}, \phi_{\rm A_1}, \phi_{\rm E})$ and its symmetry equivalent
$I_3(\theta_{\rm E}, \phi_{\rm A_2}, \phi_{\rm E})$
as well as
$I_4(\theta_{\rm E}, \phi_{\rm A_1}, \phi_{\rm A_2}, \phi_{\rm E})$.
Correlations computed based on configurations generated
using classical molecular dynamics (MD) simulations are
marked using red lines, while the corresponding
quantum path integral molecular dynamics (PIMD) data
are given by black lines.
\label{fig:34body}
}
\end{figure}

After having identified and discussed all relevant \mbox{two-,} 
three- and four-coordinate correlations in protonated acetylene, 
let us next focus on their temperature dependence and the impact 
of nuclear quantum effects.
The temperature dependence of all significant two-coordinate correlations
is presented in Fig.~\ref{fig:2body}
whereas the important three- and four-point correlations
are found in Fig.~\ref{fig:34body}.
We can clearly see that the results obtained from classical (MD) and
quantum (PIMD) simulations differ considerably
except in the limit of very high temperatures
where the classical approximation approaches
the quantum benchmark. 
As a first summary, it can be concluded that
the correlations observed for classical nuclei are always stronger
then those observed when quantum effects are included
(except for obvious artifacts at sufficiently low temperatures
where the classical approximation qualitatively fails
to describe the system). 
This indicates that quantum delocalization tends to weaken the correlations
over the whole temperature range studied here.

In the next step, we discuss the temperature evolution of the
involved correlations in detail.
We start by analyzing the classical results obtained from MD simulations.
Generally speaking all correlations increase with temperature up to a certain
point and then start to decay rapidly as temperature increases further.
This trend holds true not only for the two-body correlations but is present
in the many-body correlations as well, see Fig.~\ref{fig:34body} 
for three- and four-coordinate correlations.
At low temperatures, classical dynamics pins the molecule close to its
global minimum energy structure where it can only perform 
small-amplitude motion (mostly vibrations around that minimum energy structure).
Upon increasing the temperature, correlations strengthen
with reference to what is found at low temperatures, which can
be understood in terms of increasing the PES landscape that becomes 
available to the system.
As more thermal energy is provided, the system is able to explore
larger regions of the PES such that more complex large-amplitude 
motions can unfold from which significant correlations
are able to develop.
In the high temperature limit, the relative importance of 
potential versus kinetic energy shifts toward the latter. 
Thus, the motion of the molecule is no
longer strongly governed by the PES, but starts to be mostly a 
consequence of random thermal fluctuations.
It is, thus, expected that further increase of the temperature will
results in systematically decreasing correlations
before, eventually, the molecule must fragment 
along particular dissociation channels
which have not been considered in the present case.

%
In stark contrast to these classical results, the quantum simulations
of protonated acetylene 
provide a completely different behavior for temperatures below 100\,K,
not only in terms of the estimated correlation strength but
also in terms of the response to temperature changes. 
As can be seen in Fig.~\ref{fig:2body}, the two-body correlations 
obtained from the quantum simulations are very small and,
moreover, temperature independent below about 100\,K.
This observation also holds true for the higher-order correlations 
presented in Fig.~\ref{fig:34body}.
This combined analysis of all relevant correlations in
protonated acetylene, thus, reveals the complete absence of 
any correlations in this temperature regime~-- if nuclear quantum effects
are properly accounted for.
At the same time, the classical correlations not only increase strongly
with temperature, but also the estimated interaction information can be larger
by two orders of magnitude 
compared to the low-temperature quantum limit.

In the quantum regime, protonated acetylene is essentially in its 
quantum ground state, thus quantum fluctuations play a dominant role
whereas additional thermal activation effects are negligible.
In other words: Classical simulations completely fail to correctly 
describe the correlations of protonated acetylene in the temperature regime
below 100\,K even at the qualitatively level. 
In this context, we draw attention to the fact 
that the $I_2(r_{\rm E}, \theta_{\rm E})$ and $I_2(\phi_{\rm A_1}, \phi_{\rm A_2})$
interaction information obtained from classical MD (Fig.~\ref{fig:2body})
becomes even lower than its PIMD counterpart at these low temperatures.
This effect is due to freezing out of the involved coordinates
which is an unphysical artifact in view of the
intrinsic zero-point vibrational motion even at 0~K.

For temperatures above approximately 800\,K, however, 
protonated acetylene is reaching
the classical regime where both classical and quantum simulations
results start to be in quantitative agreement.
This is exactly that temperature regime where the
correlations, which initially increased upon heating,
hit their maxima and thereafter systematically decay
at even higher temperatures.
That qualitative change in temperature response and, thus, disruptive 
behavior of all correlations within protonated acetylene 
is observed irrespective if the nuclei are treated 
as classical or quantum point particles.
Although the classical simulations still overestimate 
the interaction information due to the neglect of quantum fluctuations,
classical MD simulations can be used safely
to study correlations of protonated acetylene in this regime. 
Finally, for temperatures in between roughly~100 and 800\,K a
quantum-classical cross-over regime is observed in which the classical
results are qualitatively in agreement with the quantum results,
but the values of the correlations are heavily overestimated. 
At variance with low temperatures, quantum fluctuations are not
dominant anymore in this regime as thermal fluctuations are
becoming increasingly more important.
For sufficiently large temperatures the quantum fluctuations
are completely overwhelmed by thermal activation effects and the
system enters the classical regime which is associated with the 
aforementioned bending of the curves at around 800\,K.

The turning points, corresponding to the temperature
at which a given correlation reaches its maximum value,
is different in the classical and quantum treatment.
It is observed that
in case of quantum simulations these maxima are shifted towards
higher temperatures by approximately 300\,K. 
This phenomenon might be viewed as a remnant of quantum 
fluctuation effects that still contribute to the correlations.

Overall these results show the ability of the correlation analysis 
based on configurational entropy presented here to unravel complex 
correlated motion within fluxional molecules such as protonated 
acetylene. 
An important finding of this study is the striking difference between 
classical and quantum correlations highlighting the destructive 
character of quantum delocalization and zero point vibration that 
essentially destroy any correlations at low temperatures 
and, in particular, in the deep quantum regime.
Even without knowing any details concerning the specific molecule,
our entropy-based correlation analysis is able to uncover that the
classical approximation to nuclear motion is entirely meaningless 
below about 100\,K, while being qualitatively correct in between 
100 and roughly 800\,K and that it finally quantitatively describes
protonated acetylene at temperatures of 1000\,K and beyond.

\section{Conclusions and Outlook}
\label{sec:conclusions}
In summary, we have outlined
a systematic and rigorous
analysis technique based on configurational entropy
that allows one to quantify temperature-dependent structural correlations,
both classical and quantum, within fluxional molecules.
This general framework, which uses concepts originating
in information theory, namely the non-parametric $k$-th
nearest-neighbor configurational entropy estimation,
can be used to decipher high-order
correlations between any desired degrees of freedom
without linearization or assumption of
any parametric model.
Applied to protonated acetylene C$_2$H$_3^+$, an archetypal
fluxional molecule, this analysis is able to unravel the
intricate impact of temperature and nuclear quantum effects
on intra-molecular structural correlations
based on $n$-coordinate (a.k.a. $n$-body or $n$-point) 
interaction information estimators up to $n=4$. 
Application to protonated acetylene shows that for our 
set of ten generalized coordinates
chosen to characterize protonated acetylene, only
three two-coordinate correlations have significant
relevance for the complex configurational coupling
in this molecule.
In addition, two symmetry-equivalent
three-coordinate correlations and a single four-coordinate correlation
feature non-negligible magnitudes in the present case.
Thus, protonated acetylene already provides a variety
of correlation phenomena that go beyond the simple
two-coordinate picture~--
as would be readily available using traditional
covariance or principal component analyses 
based on the respective correlation coefficient matrix~--
although it certainly is a small molecule.
This highlights that higher-order correlations
significantly contribute to the overall complexity of the system  
and cannot be viewed as minor corrections.

Applying the technique to configuration ensembles that have
been sampled from the classical and quantum canonical ensemble
of the nuclear skeleton reveals three distinct temperature regimes
as judged exclusively from inspecting the 
temperature-dependence of the $n$-body correlations. 
Below about 100\,K the classical description of the nuclei
provides unphysical artifacts whereas 
it works quantitatively correctly at temperatures of
1000\,K or higher.
In an intermediate regime from about 100 to 800~K, 
the quantum-to-classical cross-over occurs where classical nuclei fluctuating
due to thermal activation qualitatively mimic the
quantum-statistical description at the same temperature.

For classical nuclei, the obtained correlations 
start to increase continuously beyond roughly 100\,K 
until reaching a maximum at about 800\,K
which is the temperature of the turnover to 
a rapid entropy-driven decay due to the overriding
influence of the kinetic energy over the ordering 
effects of the potential energy.
In contrast, we overall observe strongly reduced correlations
for quantum nuclei, in particular at low temperatures
due to quantum delocalization, which is prevailing
for ground-state-dominated temperatures
up to approximately 100\,K. 
In the cross-over regime, the classical and quantum 
correlations are in qualitative agreement, albeit the 
latter are always much less pronounced than the former,
and reach quantitative agreement only in the classical 
regime at the highest temperatures where thermal 
activation dominates.  
This detailed data-intensive analysis has been
made possible due to recent developments
in machine learning approaches based on 
neural network representations of global potential energy surfaces
in high dimensions, which allows us
to describe the potential energy surface of
protonated acetylene at the essentially converged coupled cluster level.
This efficient and highly accurate approach
enables the sampling of not only the classical, 
but also of the much more demanding quantum
configurational ensemble in a statistically converged manner
even at ultra-low temperatures close to the quantum ground state. 
Combining such sampling based on neural network potentials 
with the generic entropy-based scheme for 
determining high-order correlations in large-amplitude motion 
opens the door to study many other intricate molecules as well as
inter-molecular correlations emerging in clusters or complexes.

\begin{acknowledgments}
We would like to thank Harald Forbert and Richard Beckmann for insightful discussions.
The research of R.\,T. in Bochum
was supported by a Bekker Programme scholarship
funded by the Polish National Agency for Academic Exchange
(NAWA, Narodowa Agencja Wymiany Akademickiej)~-- PPN/BEK/2018/1/00319.
C.\,S. acknowledges partial financial support from the 
Studien\-stiftung des Deutschen Volkes as well as from the
Verband der Chemischen Industrie.
This research is part of the Cluster of Excellence ``RESOLV'':
Funded by the Deutsche Forschungsgemeinschaft
(DFG, German Research Foundation) under Germany's
Excellence Strategy~-- EXC 2033~-- 390677874.
The computational resources were provided by HPC@ZEMOS,
HPC-RESOLV, and BOVILAB@RUB.
\end{acknowledgments}

\section*{References}

%

\end{document}